# Systematic reduction of Hyperspectral Images for high-throughput Plastic Characterization

Mahdiyeh Ghaffari[1]*, Mickey C. J. Lukkien[1], Nematollah Omidikia[1], Gerjen H. Tinnevelt[1], Marcel C. P. van Eijk[2], Jeroen J. Jansen [1]*

[1]Radboud University, Institute for Molecules and Materials, Analytical Chemistry, P.O. Box 9010, 6500 GL Nijmegen, the Netherlands

[2]National Test Centre Circular Plastics (NTCP), Duitslanddreef 7, 8447 SE Heerenveen, the Netherlands

**Abstract**

Hyperspectral Imaging (HSI) combines microscopy and spectroscopy to assess the spatial distribution of spectroscopically active compounds in objects, and has diverse applications in food quality control, pharmaceutical processes, and waste sorting. However, due to the large size of HSI datasets, it can be challenging to analyze and store them within a reasonable digital infrastructure, especially in waste sorting where speed and data storage resources are limited. Additionally, as with most spectroscopic data, there is significant redundancy, making pixel and variable selection crucial for retaining chemical information. Recent high-tech developments in chemometrics enable automated and evidence-based data reduction, which can substantially enhance the speed and performance of Non-Negative Matrix Factorization (NMF), a widely used algorithm for chemical resolution of HSI data. By recovering the pure contribution maps and spectral profiles of distributed compounds, NMF can provide evidence-based sorting decisions for efficient waste management. To improve the quality and efficiency of data analysis on hyperspectral imaging (HSI) data, we apply a convex-hull method to select essential pixels and wavelengths and remove uninformative and redundant information. This process minimizes computational strain and effectively eliminates highly mixed pixels. By reducing data redundancy, data investigation and analysis become more straightforward, as demonstrated in both simulated and real HSI data for plastic sorting.

## 1. Introduction

Hyperspectral Imaging (HSI) is a valuable non-destructive technology for online monitoring and high throughput screening, with huge potential to be automated [1-4]. HSI , specifically Near InfraRed (NIR), is both fast and non-destructive, which makes it excellently suitable to perform real-time "online" in identification of polymers [5, 6]. The combination of HSI as a powerful tool for the optical sorting of plastics, and discriminant analysis, allows automatic sorting in real time under industrial conditions [7] into new well-characterized circular feedstock to replace virgin polymers. The high throughput required to process sufficient material of the enormous amount of plastic waste puts strong demands on the computational time and storage resources available to support the choice for sorting every waste element into an adequate stream: data analysis, processing, and decision-making should be as fast as possible [8]. An early implementation of NIR hyperspectral imaging-based cascade detection, exemplified by industry leaders like Steinert, Tomra, and Pellend, was already able to specifically sort post-consumer plastic packaging waste. Innovations in the development of such technology have been continuously ongoing [9-11].

The considerable promise of HSI, with its challenging combinations of multivariate spectral and spatially resolved information, have made multivariate methods essential to extract hidden chemical information [12]. Chemometrics translates abstract spectroscopic fingerprints into throughput plastic species compositions of each waste element plays an indispensable role in the analysis of plastics during the recycling process. Chemometrics has proven invaluable, by variable selection [13], Multivariate Curve Resolution (MCR)[14], and classification algorithms to mix chemical domain knowledge with data-driven machine learning. It plays a vital role in tackling the complexities presented by challenging objects, particularly multilayers, as it enables the analysis and interpretation of the intricate spectral information embedded within these composite structures. However, the size of Hyperspectral data sets in both the spatial and spectral dimensions may considerably limit the speed of information extraction from such data. The Spectral and spatial redundancy of HSI data however enables a considerable data reduction to reduce resources sufficiently for high-throughput applications in the circular economy [15, 16].

Non-negative matrix factorization (NMF) with the inherent non-negativity property originates in signal/image processing with some reports in analytical chemistry yet is largely mathematically equivalent to Multivariate Curve Resolution (MCR-ALS) [17, 18] that is a cornerstone of chemometrics. The NMF algorithm has been evaluated in chemistry e.g. for the resolution of overlapped GC-MS spectra, as well as for the deconvolution of GC × GC data set [19]. Besides, NMF has been conducted to classify complex mixtures based on the extracted feature and to analyze time-resolved optical waveguide absorption spectroscopy data [16]. NMF recovers pure spectra and their concentration contribution maps of HSI Raman images [20]. Another study applied an NMF filter to triboluminescence (TL) data traces of active pharmaceutical ingredients [21], leading to simultaneously recovering both photon arrival times and the instrument impulse response function, i.e. is of considerable value to recover chemical information from HSI data. Zushi proposed an NMF-based spectral deconvolution approach, coupled with a web platform GC mixture [22], that utilizes a faster multiplicative update method instead of the traditional projection step. This approach is highly advantageous for analyzing large mass spectrometry imaging datasets due to its improved speed [23].

Monolayer materials composed of a single polymer species or multilayer materials made up of several polymers are common in plastics. With the help of a single spectral profile, it is possible to identify the polymer composition of monolayer objects [24]. When it comes to identifying the polymer composition of waste streams, the objects encountered are often much more complex than monolayer materials. These objects can be multilayered or composed of multiple polymer species, or even coated with labels made of different polymers. Furthermore, an object may be made up of known polymers with an unknown composition. In the case of multilayer plastics, the type and ratio of plastics used can also vary significantly, such as 70/30 or 50/50. This adds significant complexity to the classification process, which is where the importance of data unmixing comes into play.

Data unmixing is a valuable tool that can handle these issues effectively. However, when sorting the objects, the decision to translate a model observation into a specific engineered materials stream may only be made on a small subset of the complete object. Therefore, reducing the size of the data using convex hull is an excellent innovation in curve resolution that can reduce the data to less than one percent of the original data. Handling large HSIs, data compression/size reduction coupled with rapid data analysis tools has top priority. So, unnecessary pixels (data rows) and wavelengths (data columns) should be jointly removed and the rest are essential to characterize. For the study in this research, removing all unnecessary information and the reduced data will be subjected to further decomposition, NMF. Simulated and real experimental data are exemplified.

## 2. materials and methods

### 2.1. A brief description of Non-Negative Matrix Factorization

Nonnegative Matrix Factorization (NMF) deconvolutes a matrix, $\mathbf{R}_{IJ \times K}$, into the product of matrices $\mathbf{W}_{IJ \times n}$ and $\mathbf{H}^T_{n \times K}$ with an intrinsic non-negativity property as a minimal constraint. *I* and *J* represent the number of spatial pixels, while *K* represents the number of variables in the case of a hyperspectral image.

$$\mathbf{R}_{IJ \times K} = \mathbf{W}_{IJ \times n} \mathbf{H}^T_{n \times K} + \mathbf{E}_{IJ \times K} \qquad (1)$$

To this, formalize NMF optimizes the following cost function:

$$\min_{(\mathbf{W} \geq 0, \mathbf{H} \geq 0)} \left\| \mathbf{R}_{IJ \times K} - \mathbf{W}_{IJ \times n} \mathbf{H}^T_{n \times K} \right\|_F \qquad (2)$$

Different algorithms are introduced in the literature to calculate component matrices, The multiplicative update rule introduced by Le ad Seung is simple to implement[6]. The updating parts are:

$$\boldsymbol{H}_{K \times n} \leftarrow \boldsymbol{H}_{K \times n} \frac{(\mathbf{W}^T_{n \times IJ} \mathbf{R}_{IJ \times K})}{(\mathbf{W}^T_{n \times IJ} \mathbf{W}_{IJ \times n} \mathbf{H}^T_{n \times K})} \qquad (3)$$

and

$$\mathbf{W}_{IJ \times n} \leftarrow \mathbf{W}_{IJ \times n} \frac{(\mathbf{R}_{IJ \times K} \mathbf{H}^T_{n \times K})}{(\mathbf{W}_{IJ \times n} \mathbf{H}^T_{n \times K} \boldsymbol{H}_{K \times n})} \qquad (4)$$

Equations 3 and 4 denote element-wise multiplications. These update rules preserve the non-negativity of $\mathbf{W}_{IJ \times n}$ and $\boldsymbol{H}_{K \times n}$ where $\mathbf{R}_{IJ \times K}$ is element-wise non-negative.

## 2.2. Approach

For any bilinear data set, a minimum number of rows and columns carry the most informative and independent part of the data [25]. Consequently, in HSIs, essential pixels, and essential wavelengths are necessary to extract the pure contribution maps and spectral profiles of all components. The main steps of this approach for plastic characterization are visualized in Figure 1 and summarized as:

1) *Object detection:* Recording a first-order spectrum with $K$ channels for every pixel in an $I$ by $J$ scene into a data cube, $\underline{\tilde{\mathbf{R}}}_{I \times J \times K}$. To analyze this cube for plastic characterization, object detection is needed. In this work the corresponding pixels of each object were detected by correlation growing algorithm. The results of object detection on $\underline{\tilde{\mathbf{R}}}_{I \times J \times K}$ are several cubes (as many as the number of objects) and each of them contains information about one of the objects. Object detection decomposes $\underline{\tilde{\mathbf{R}}}_{I \times J \times K}$ to some cubes.

2) *Essential information extraction:* This step will start by unfolding each data cube (resulted from previous step) to a matrix ($\mathbf{R}_{IJ \times K}$) with each row a pixel for further analysis. Calculating the most informative pixels (Essential Spectral Pixels, ESPs) and variable/wavelengths (Essential Spatial Variables, ESVs) for each object, which are based on the convexity property in the normalized abstract row and column spaces, is the next step. ESP/ESVs are the smallest set of points needed to generate the whole data in a convex way in the abstract score space. Once ESP/ESVs are identified for all objects separately, all other measured pixels are removed and the reduced data is moved to the next step. Left/right eigenvectors of $\mathbf{R}_{IJ \times K}$ using SVD, $\mathbf{R}_{IJ \times K} = \mathbf{U}_{IJ,n}\, \mathbf{D}_{n,n}\, \mathbf{V}_{n,K}^{T} + \mathbf{E}_{IJ,K}$ can be calculate, where $\mathbf{U}_{IJ,n}$ and $\mathbf{V}_{n,K}^{T}$ are the left and right eigenvectors, respectively. $\mathbf{D}_{n,n}$ and $\mathbf{E}_{IJ,K}$ contains singular values and residuals, individually. In addition, $n$ is the number of factors. The number of factors is set up to five. Because most of the objects contain less than five types of polymers/materials. "Convhulln" as a MATLAB function can calculate the convex set of $\mathbf{U}_{IJ,n}$ and $\mathbf{V}_{n,K}^{T}$ and explore the ESPs/ESVs. After join selection of essential pixels and essential variables, $\mathbf{R}_{IJ \times K}$ turn into $\boldsymbol{R}_{p_{ESP} \times K_{ESV}}$. This step is explained in detail in the previous works [25].

3) *Data decomposition:* The reduced data sets for all objects, $\boldsymbol{R}_{ESPs \times ESVs}$ can be analyzed by NMF in parallel with multiplicative updates, to calculate the reduced concentration contribution maps, $\boldsymbol{W}_{ESPs \times n}^{r}$ and reduced spectral profiles $\boldsymbol{H}_{ESVs \times n}^{r}$ using Eq. 3 and 4. Using least square, full concentration contribution maps, $\boldsymbol{W}_{IJ \times n}$, and full spectral profiles, $\boldsymbol{H}_{K \times n}$ can be produced from reduced versions $W^r_{ESPs \times n}$ and $H^r_{ESVs \times n}$ through eqs 6 and 7. This step needs $\mathbf{R}_{ESPs \times K}$ and $\mathbf{R}_{IJ \times ESVs}$ which are one-mode reduced data in the row and column direction respectively. Finally, it is easy to reshape the columns of $\boldsymbol{W}_{IJ \times n}$ to generate the full concentration contribution maps.

$$\boldsymbol{W}_{IJ \times n} = \boldsymbol{R}_{IJ \times ESVs} * \text{pinv}(\boldsymbol{H}^r_{ESVs \times n}) \qquad (6)$$

$$\boldsymbol{H}_{K \times n} = pinv(\boldsymbol{W}^r_{ESPs \times n}) * \mathbf{R}_{ESPs \times K} \qquad (7)$$

4) *Decision making:* The matrix $\boldsymbol{H}_{K \times n}$ consists of pure spectral profiles of the components, which can be utilized for qualification by comparing them to a reference library. However, in some cases, $\boldsymbol{W}_{IJ \times n}$ has complementary information for identification. $\boldsymbol{W}_{IJ \times n}$, contains characteristic information about the composition of unknown objects. The sum of the squares of the elements in each column of $\mathbf{W}_{IJ \times n}$ represents the variance of the signal contributed by each polymer in an unknown object. This variance can be used to differentiate between mono-material and multi-material objects using a statistical F-test, which is commonly employed in statistics to compare the standard deviations of two populations. To conduct the F-test, the variance of all columns in $\mathbf{W}_{IJ \times n}$ is calculated in the first step, and then each value is divided by the noise variance. A significance level of $P < 0.05$ is used to determine whether the variance is statistically significant. If the object contains only one layer of material, only one type of polymer will pass the F-test. However, if it is a multilayer object, multiple types of polymer will pass the test. Finally, it should be emphasized that all of the computation and reported times in this work are based on a laptop (Intel(R) Core(TM) i7-10850H CPU @ 2.70GHz) which for real industrial purposes can be dramatically improved by using better computers. The utilization of this algorithm enables the exploitation of the computational capabilities of a standard computer system, thereby eliminating the reliance on specialized hardware such as GPUs or high-performance computing clusters.

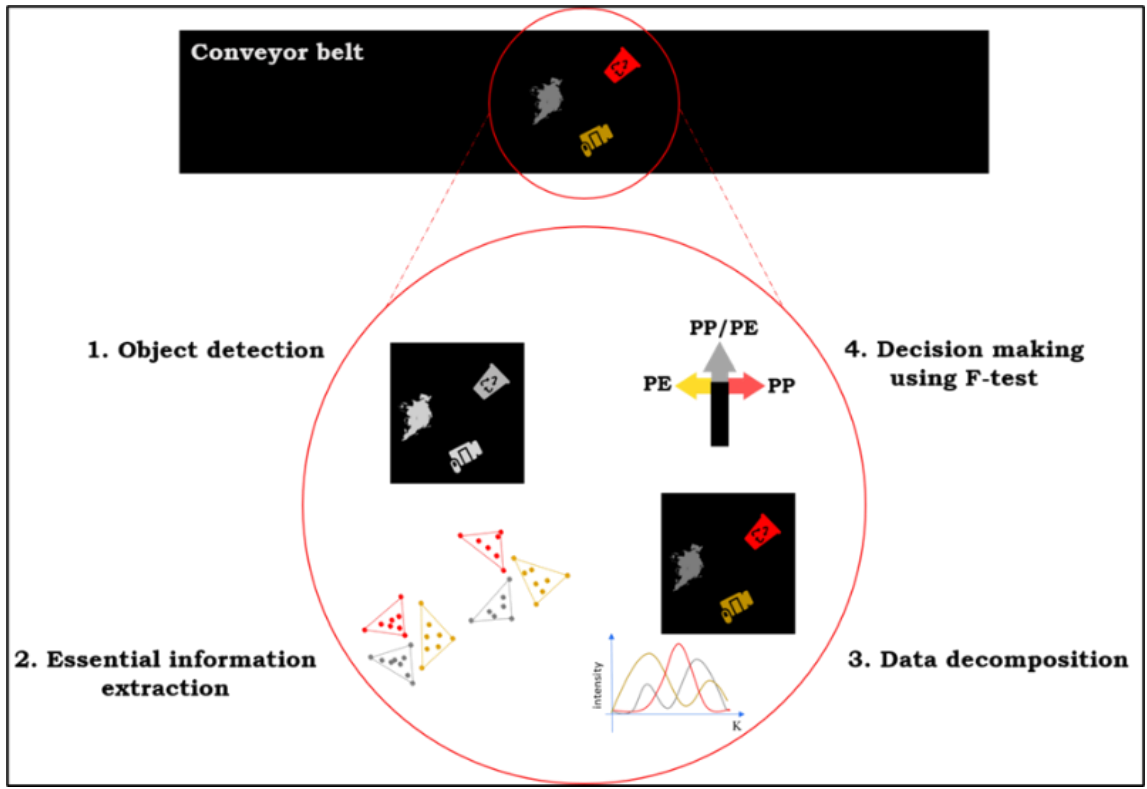

**Fig. 1.** A Graphical illustration of the approach based on the essential pixel/wavelength selection and NMF. The hyperspectral data with a 3D structure, needs object detection first. Then the selected parts of the data which correspond to the objects unfold to matrices to calculate the ESPs/ESVs. Finally, the reduced data sets are analyzed by NMF and full contribution maps and spectral profiles are retrieved. In the last step, F-test will help for decision making to monolayer/multilayer plastic sorting.

## 3. Data description

### 3.1. Simulated data set

A hyperspectral image was simulated to visualize the effect of essential pixels and variables selection using convex polytope and further unmixing by NMF. The concentration contribution maps (re-folded concentration profiles) and pure spectral profiles are shown in Figure 2. The simulated hyperspectral data set is of dimensions 253 × 186 pixels by 141 variables and the unfolded two-way data matrix of 47058 pixels and 141 pseudo-spectral channels. Despite the simplicity of the simulation, it should be noted that care was taken to avoid the pure pixels or selective spectral channels, this corresponds to a non-trivial situation for NMF analysis. For this purpose, small random numbers were added to the pure contribution maps. In this case, eight objects are on the hypothetical conveyor belt which are made of polypropylene (PP), polyethylene (PP), and polyethylene terephthalate (PET). Five and three objects are monolayers and multilayers, respectively. Figure 2 presents the pure contribution maps and spectral profiles of PP, PE, and PET.

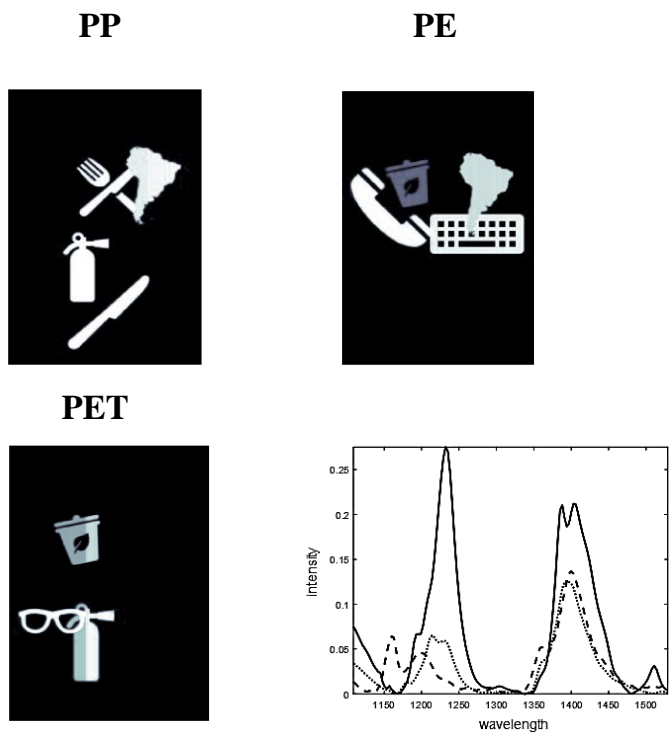

**Fig. 2.** The three components simulated the HSI data set. Concentration distribution maps and pure spectral profiles are presented.

## 3.2. Experimental Hyperspectral Images of Plastics

A collection of monolayer and multilayer objects made from PP (polypropylene), PS (polystyrene), PET (polyethylene terephthalate), and polypropylene on top of polyethylene (multilayer), all with known compositions, were gathered from packaging waste. The characteristic information of these objects was known in advance. To collect data, the objects were randomly placed on a conveyor belt, and two HSI measurements were recorded in the 900-1700 nm range. The resulting data sets are presented in Figure 3.

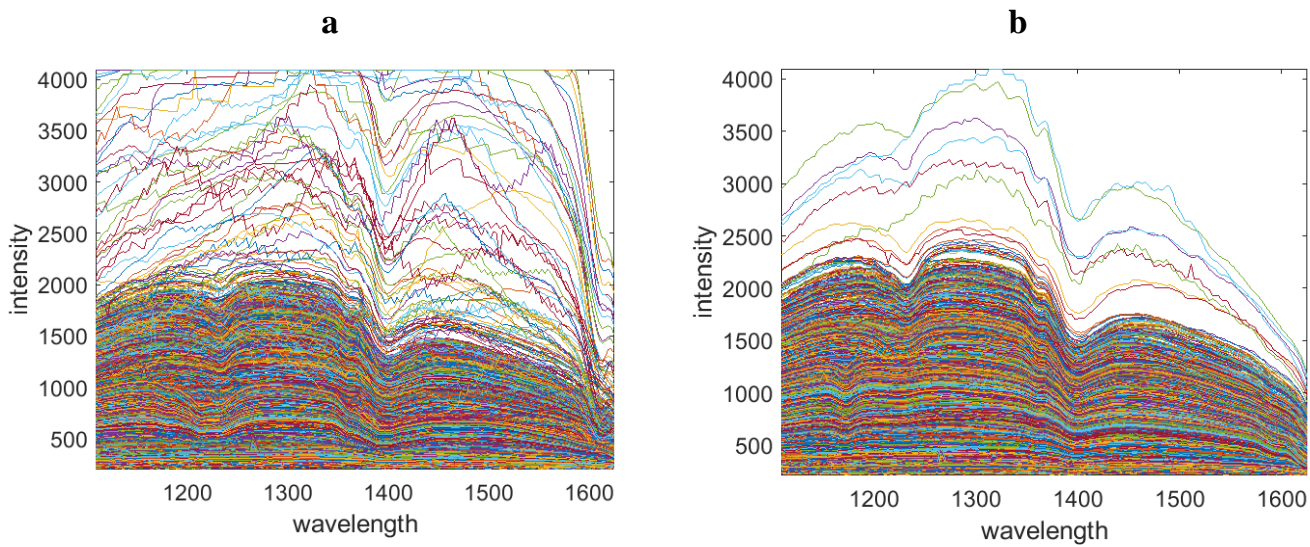

**Fig. 3.** The experimental cases are visualized in a and b.

## 4. Result and Discussions

To illustrate the effect of essential spectral pixels (ESPs) and/or essential spectral variables (ESVs) selection of the plastic sorting using NMF, the results obtained on the simulation and real HSI data set are discussed in detail.

The procedure starts with object detection for the simulated case as it does not need any data per-processing. Then, the ESPs and ESVs should be selected using a convex hull on the normalized abstract spaces of the data set for each object. First five Principal Components were used to generate the abstract spaces. The selected essential pixels and wavelengths are presented in Figures 4a and 4b. In Figure 4, the left and right panels contain the mean image and the spectral data of the simulated case. The selected ESPs and ESVs are shown by white crosses and red lines. The raw data matrix of the simulated case is composed of 57078 rows (pixels) and 173 variables were reduced to eight matrices by object detection. For each object, two or three ESPs and a few ESVs are selected as it is shown in Figure 4a-b. In total for the whole data, only 0.04% data variance remains. It means only 0.04 percent of the data is essential/enough to analyze.

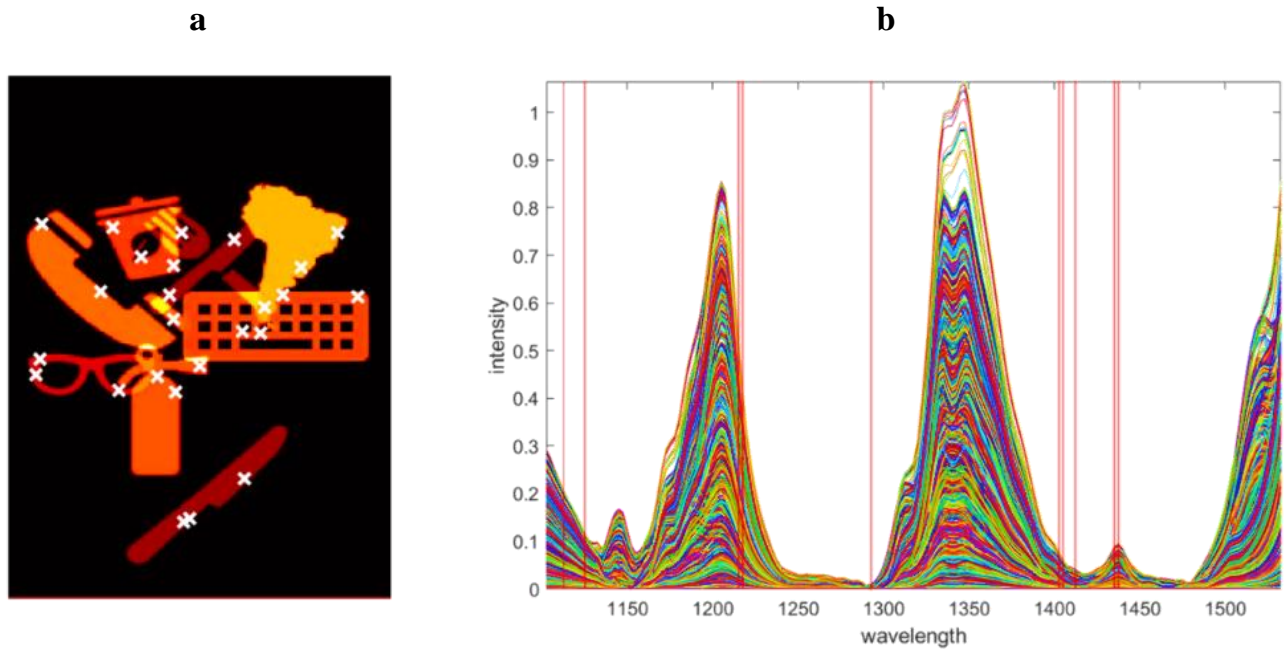

**Fig. 4.** The selected ESPs and ESVs for the simulation case are indicated as white crosses and red lines, respectively.

Later, the reduced data sets were analyzed by NMF. The results of NMF are called **W** and **H**. To make a better visualization of the full contribution maps, **W** changed to binary matrices and visualized in Figure 5. The total calculation time for NMF analysis of the reduced data sets was nearly 0.001 sec, 1% of the time required for the same calculations on all pixels and wavelengths.

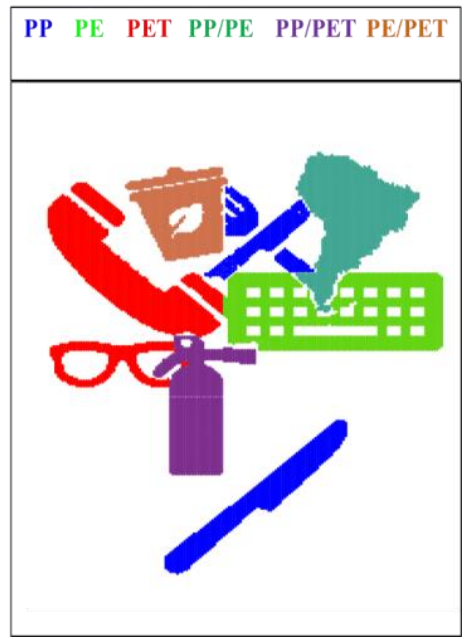

**Fig. 5.** The full contribution maps of all components in simulated data are visualized. Each color is used for special composition. Blue, green, red, aqua, and purple, brown, are used for, PP, PE, PET, PP/PE (multilayer), PP/PET, and PE/PET, respectively.

In addition, strategic data pre-processing methods are essential prior to data analysis to linearize the data, remove artifacts and thereby optimally align the data and the NMF with the Beer–Lambert–Bouguer law. Figure 6 presents the effect of each pre-processing step on the shape of the spectral data for an example object. Figure 6a is the recorded raw HSI of an optional object. Some spectral profiles which reflected the light can be removed first. These profiles can be recognized by the slope of the spectral profiles in the range of 1100-1500 nm. In this wavelength range the slope of the profiles are zero (all saturated).

Figure 6b presents the normalized data after removing the saturated spectral profiles. The next step (6c) is the treated data after baseline correction of the using asymmetric least squares on the minus logarithm of the data set. Finally, the data set is reconstructed by a few principal components for denoising purposes. The corrected data is presented in Figure 6d.

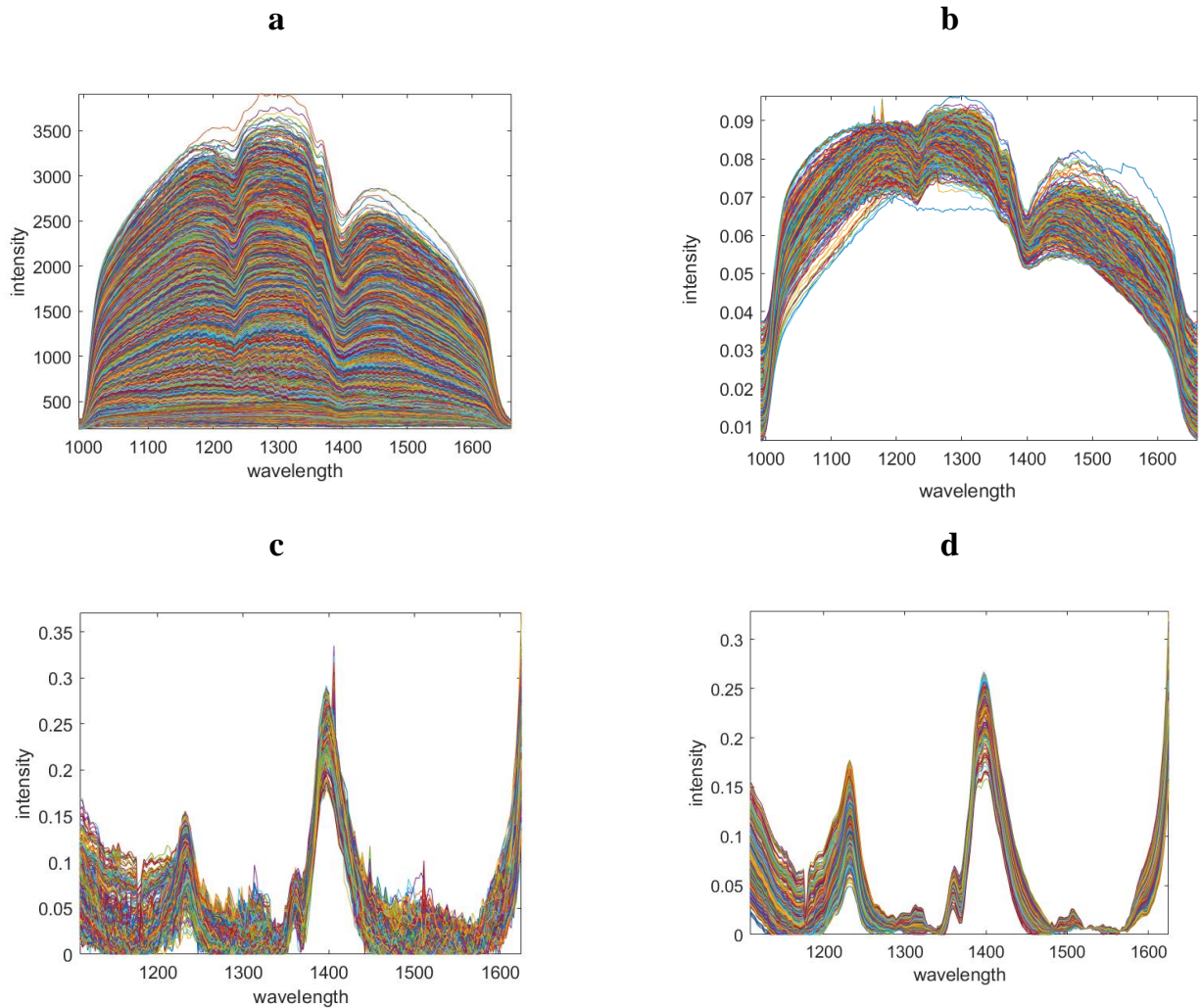

**Fig. 6.** Depicts the data set corresponding to the first unknown object, visualized after undergoing various pre-treatment steps. These steps include object detection and removal of pixels affected by signal saturation, baseline correction of the negative logarithm of the data, and denoising using Singular Value Decomposition.

Once the data has been preprocessed, it undergoes object detection, as detailed in the theory section. This process yields smaller data sets, each one corresponding to a single object and containing relevant information. In the current scenario, object detection yields five data sets (corresponding to the five objects on the conveyor belt). The crux of the proposed algorithm lies in selecting the appropriate ESPs and ESVs based on the data's inner polygon in abstract spaces. Figures 7 and 8 illustrate the chosen ESPs and ESVs for both real cases, denoted by white crosses and red vertical lines. Finally, the size of the data is decreased to $48 \times 12$ and $149 \times 45$, from the original dimensions of $144000 \times 173$.

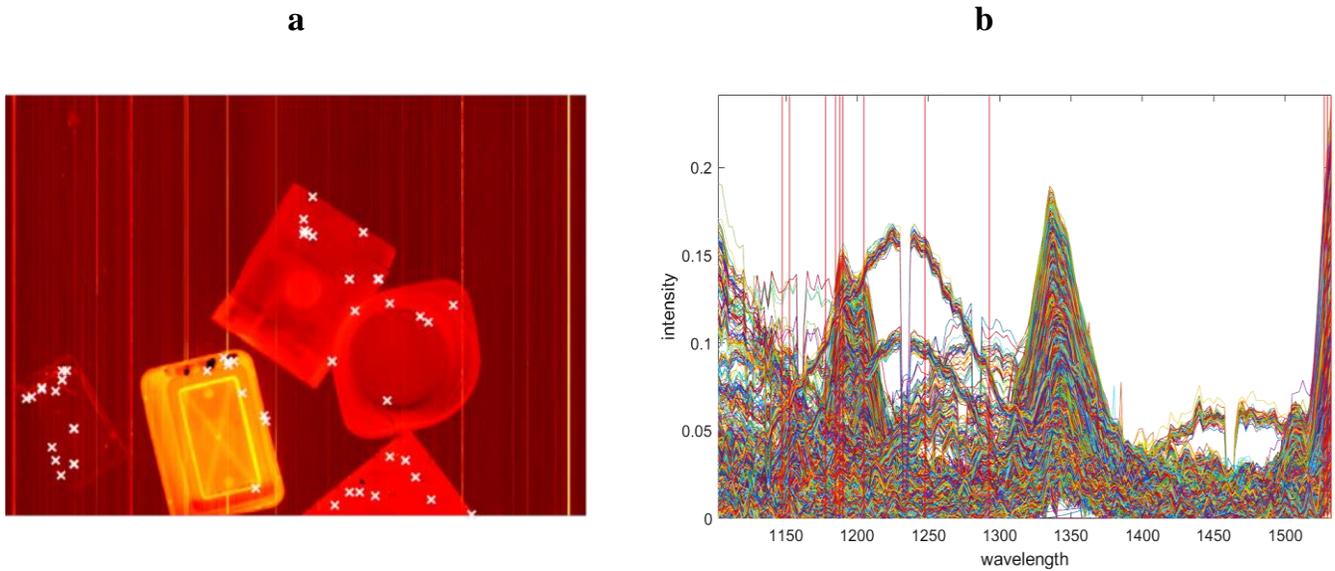

**Fig. 7.** Depicts the mean image and spectral profile of the first experimental case, presented in (a) and (b) respectively. The chosen ESPs and ESVs are denoted by white crosses and red lines.

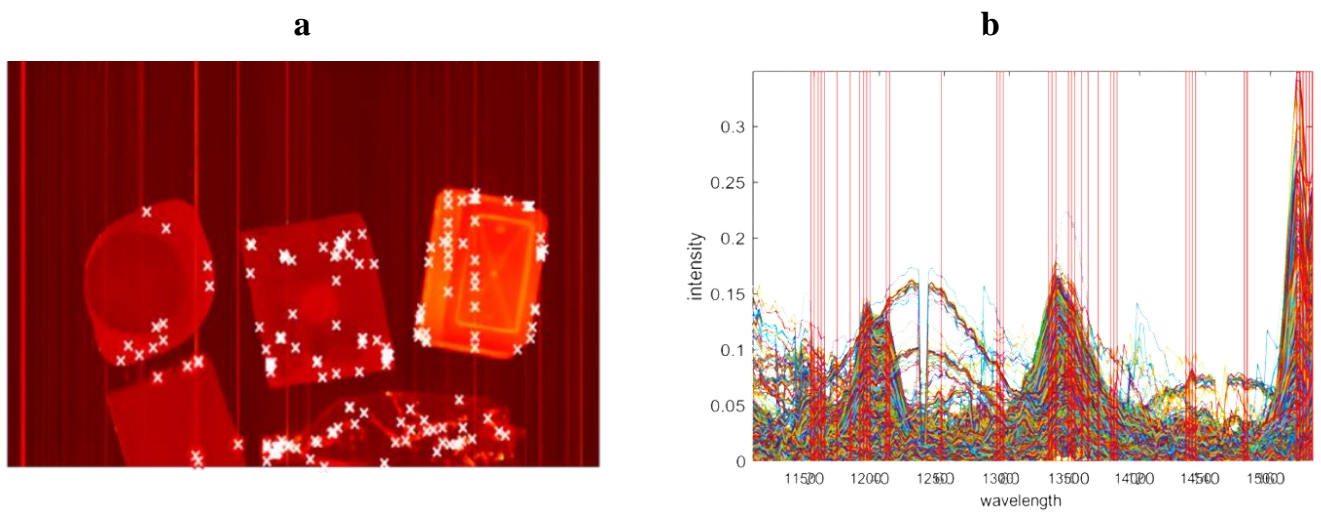

**Fig. 8.** Depicts the mean image and spectral profile of the second experimental case, presented in (a) and (b) respectively. The chosen ESPs and ESVs are denoted by white crosses and red lines.

The reduced data sets were analyzed by NMF. The final resulting maps which identified the composition of each object are presented in Figure 9. To generate these single contribution maps, one least square with a non-negative least square was necessary which was followed by binary maps construction. In the resulting images, each color is used for special composition as is indicated on the top of the figure as well. For comparison, the full data sets were analyzed by NMF as well which took almost 220 seconds. However, NMF for the reduced data sets took 0.003 second.

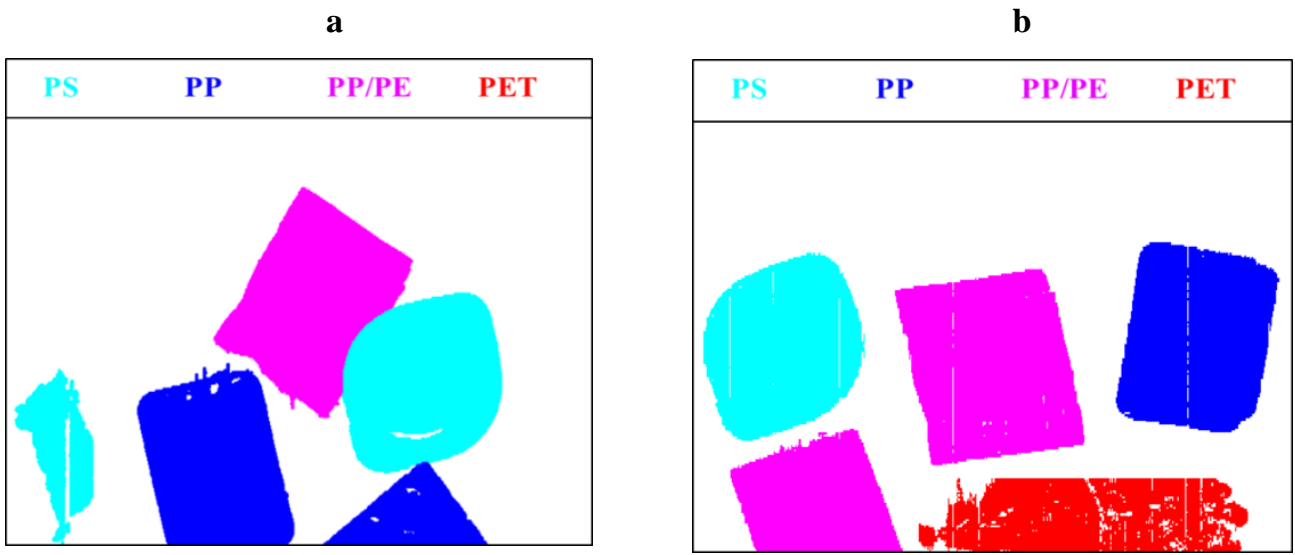

**Fig. 9.** The full contribution maps of all components in both experimental data sets are visualized in a and b. Each color is used for special composition. Cyan, blue, purple, and red, are used for PS, PP, PP/PE (multilayer), and PET, respectively.

## 5. Conclusion

In this contribution, a simple and efficient procedure for the selection of essential pixels and wavelengths in HSI data sets is introduced with a high potential for application in industrial plastic sorting. Nowadays, HSIs are used in plastic sorting with the aim of plastic-type identification. So coupling data size reduction (which seems logical in big HIS analysis) with a fast algorithm is promising. The benefit of data reduction is not just for the analysis part. However, the reward can be used in the data recording scheme. Considering the wide NIR domain, recording a few wavenumbers rather than the whole scope can be dramatically advantageous. On the other hand, NMF is suggested to decompose reduced data in plastic sorting. This procedure can be induced to analyze data from complementary domains, like remote sensing, etc. However, plastic sorting is a case in that we explained the procedure based on it.

**Acknowledgment**

This project is co-funded by TKI-E&I with the supplementary grant 'TKI- Toeslag' for Top consortia for Knowledge and Innovation (TKI's) of the Ministry of Economic Affairs and Climate Policy. We thank all partners in the project "Towards improved circularity of polyolefin-based packaging", managed by ISPT and DPI in the Netherlands. It was partly funded by the Perfect Sorting Consortium, a consortium that develops AI technology to enable the intended use of sorting for the recycling of packaging material. The members are NTCP, Danone, Colgate-Palmolive, Ferrero, LVMH, Mars, Michelin, Nestlé, Procter & Gamble, PepsiCo, Ghent University, and Radboud University.

**Data availability**

The datasets used and/or analysed during the current study available from the corresponding author on reasonable request.